# X-ray optical systems: from metrology to Point Spread Function

Daniele Spiga[1§] and Lorenzo Raimondi[2]

[1]INAF / Brera Astronomical Observatory, Via Bianchi 46, 23807 Merate (Italy)
[2]Elettra Sincrotrone Trieste SCpA, Area Science Park, 34149 Basovizza (Italy)

**ABSTRACT**

One of the problems often encountered in X-ray mirror manufacturing is setting proper manufacturing tolerances to guarantee an angular resolution - often expressed in terms of Point Spread Function (PSF) - as needed by the specific science goal. To do this, we need an accurate metrological apparatus, covering a very broad range of spatial frequencies, and an affordable method to compute the PSF from the metrology dataset. In the past years, a wealth of methods, based on either geometrical optics or the perturbation theory in smooth surface limit, have been proposed to respectively treat long-period profile errors or high-frequency surface roughness. However, the separation between these spectral ranges is difficult do define exactly, and it is also unclear how to affordably combine the PSFs, computed with different methods in different spectral ranges, into a PSF expectation at a given X-ray energy. For this reason, we have proposed a method entirely based on the Huygens-Fresnel principle to compute the diffracted field of real Wolter-I optics, including measured defects over a wide range of spatial frequencies. Owing to the shallow angles at play, the computation can be simplified limiting the computation to the longitudinal profiles, neglecting completely the effect of roundness errors. Other authors had already proposed similar approaches in the past, but only in far-field approximation, therefore they could not be applied to the case of Wolter-I optics, in which two reflections occur in sequence within a short range. The method we suggest is versatile, as it can be applied to multiple reflection systems, at any X-ray energy, and regardless of the nominal shape of the mirrors in the optical system. The method has been implemented in the WISE code, successfully used to explain the measured PSFs of multilayer-coated optics for astronomic use, and of a K-B optical system in use at the FERMI free electron laser.

## 1. INTRODUCTION: THE PROBLEM OF PSF COMPUTATION IN X-RAY MIRRORS

The prediction of the optical performances of X-ray mirrors is a crucial problem in the treatment of metrological data. An accurate simulation of the imaging quality allows us to determine if an optical system is suitable for the specific application, and conversely to establish manufacturing tolerances. The optical quality is usually expressed quantitatively via the PSF (*Point Spread Function*), i.e., the annular integral of the intensity around the center of the focal spot, divided by the annulus radial amplitude and normalized to the radiation intensity collected by the mirror aperture. The PSF calculation from metrology (including profiles measured e.g., with LTP[1], and roughness measured with methods like PSI, Phase Shift Interferometry, or AFM, Atomic Force Microscopy) is therefore a fundamental task to check the optical performances for both astronomical and Synchrotron/Free Electron Laser (FEL) imaging applications.

In X-ray astronomical mirrors, angular resolutions HEW (*Half Energy Width*, the angular diameter enclosing 50% of the focused rays) of a few arcsec are required to avoid source confusion in astronomical images. At the same time, optics for X-ray telescopes require a large effective area and need to be operated in space, so they are typically manufactured nesting several grazing incidence mirrors, each of them with Wolter-I[2] or polynomial[3] longitudinal profile, into a densely packed assembly. The mirror walls have to be kept as thin as possible (a few tenth mm) to ensure high filling of the telescope aperture, but keeping the mass within acceptable limits for the launch. In turn, thin mirrors are prone to deform, which goes at the expense of the angular resolution. In addition, the large number of mirrors to be manufactured entails an industrial production process, in which the surface of each individual mirror cannot undergo a dedicated polishing process. For this reason, a tolerable level of profile errors and surface roughness, strictly depending on the scientific requirement on the HEW, shall be established prior to manufacturing. For example, the ATHENA X-ray telescope[4], selected for the L2 slot in ESA's Cosmic Vision 2015–25 with a launch foreseen in 2028, requests an effective area of 2 m$^2$ and a HEW < 5 arcsec at 1 keV. This challenging target will require not only a dedicated advanced technology, but also dedicated metrology tools and a reliable method to consistently predict the PSF from the metrology dataset.

---

[§] contact author: Daniele Spiga, phone +39-039-5971027, email: daniele.spiga@brera.inaf.it

A similar need is encountered in X-ray mirrors in use at Synchrotron of FEL beamlines[5] to focus or collimate X-ray beams, but in this case high effective area is not necessary because the X-ray beam is intense and directional, so only a few mirrors are to be manufactured. In contrast, a focusing accuracy to within a few microns (FWHM, Full Width Half Maximum) is necessary for most applications. Fortunately, these mirrors can also have a relevant weight because they have not to be launched to space, so the manufacturer can concentrate on improving the focusing performances at their best, possibly making use of benders/actuators for an in-situ shape optimization. Nevertheless, aiming to determine the surface finishing tolerances and optimize the mirror profile, also in this case the PSF has to be carefully computed.

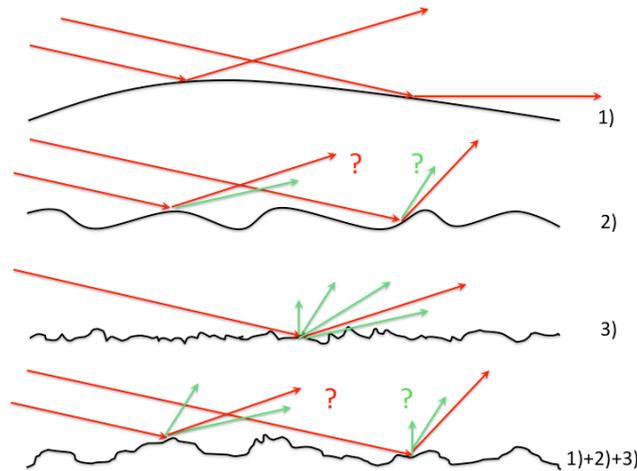

Fig. 1: X-ray mirror defects are classically decomposed in bands of different spatial wavelengths and classified on the basis of the behavior they exhibit when reflecting/scattering the radiation. 1) Long-period deformations are easily treated via geometrical optics. 2) Mid-frequency errors are of uncertain treatment. 3) The effect of microroughness, in the smooth-surface limit, can be effectively computed using the first order scattering theory. The general case includes all 3 sources of profile errors and can be difficult to treat self-consistently.

Many authors have so far studied the problem of PSF calculation. The difficult aspect is that, at a first glance, X-rays exhibit different behaviors (Fig.1), depending on the mirror defect size. Consequently, we are used to distinguish between different spectral regimes. The effect of *long-period wavelengths* on the PSF can be studied using geometrical optics, for example via ray-tracing programs. Very short wavelengths compose the realm of *microroughness*, in which the geometrical optics cannot be used, but the wave nature of X-rays have to be explicitly taken into account and give rise to the X-ray scattering (XRS). The scattering effect results in an increasing PSF degradation with the X-ray energy; the scattering of a stochastically rough surface can be computed from the statistical description of the roughness expressed in terms of PSD (*Power Spectral Density*[6]), which for highly-polished surfaces can be well approximated by a power-law of the spatial frequency[7]. If the smooth-surface condition is fulfilled, we usually apply the 1$^{st}$ order scattering theory[8], in which the scattering distribution is proportional to the surface PSD[9],[10],[11].

However, the separation between these two behaviors is not immediate to set. An attempt was made in 2005 by Aschenbach, who proposed[12] the smooth-surface limit applied to each Fourier component as a criterion to classify it as roughness (to be treated with the 1$^{st}$ order theory) or figure (to be studied with ray-tracing methods). Despite being credited for setting physical bases to a problem previously solved "intuitively", this approach has some drawbacks:

1) There is no reason to believe that the transition ray-tracing/scattering is abrupt: in fact, between the figure and roughness regimes there is a sort of "no man's land" of spatial wavelengths in the centimeter/millimeter range ("*mid-frequencies*"), in which the scattering theory cannot usually be applied, but also the utilization of geometrical optics is doubtful (Fig. 1). In other words, geometrical optics and the scattering theory represent two *asymptotical* regimes, but are unable to describe the intermediate situations. As we will see in Sect. 3.5, the PSF of mid-frequencies can be quite complicated: hence, if mid-frequencies are present in an X-ray mirror profile, the PSF computation can be affected by a relevant degree of uncertainty.

2) The "rms of a single harmonic" makes sense only if the power spectrum is *discrete*. If the PSD is a continuous function of the spatial frequency $\nu$ (like in a power-law model), the rms of *one* harmonic is *zero*, while the rms in a frequency band [$\nu$ and $\nu+\Delta\nu$] makes a physical sense. In real cases, the profile has a finite length $L$, so $\nu$ can be *known* at minimum steps $\Delta\nu_{min} = 1/L$. This would make the separation roughness/figure a function of the mirror length (or the instrumental scan length), whereas it ought to be only a function of the surface finishing.
3) It is dangerous to apply physical models out of their domain of validity. For example, a ray-tracing routine applied to a profile including roughness defects often returns visibly wrong results. But also the application of the first-order scattering theory at frequencies near $1/L$ is often incorrect.

Finally, we mention that scattering theories exist to overcome the smooth-surface limit[13],[14], but the theory was always limited to the case of stochastically rough surfaces, and seems not to have been extended to mirrors including deterministic profile errors. However, a question still remains: even assuming that we are able to locate the boundary frequency geometry/roughness, and that we are able to correctly treat the mid-frequency range, when we have separately computed the PSFs for the 3 regimes shown in Fig. 1, *how shall we combine them together to return the final PSF*? Even if a convolution might seem a natural answer, it can be proven to be in general incorrect[15].

In this work we will provide a general and self-consistent method to compute the PSF from a profile and roughness PSD characterization of a mirror, without any assumption excepting that X-rays impinge at a shallow grazing angle, $\alpha_0$. This method, which we already presented in some previous SPIE papers[16],[17],[18], is based on the Huygens-Fresnel principle, which can be applied *without restrictions*. In this way, the old problem of setting boundary frequencies between figure/waviness/roughness, with all the aforementioned difficulties, *does no longer need to be solved*!

In Sect. 2 we see that in grazing incidence the Huygens-Fresnel principle application is simplified, because the computation can be performed in only one dimension, i.e., in the mirror longitudinal direction. In this way, we can derive an integral formula (Sect. 3) to calculate the PSF of a mirror with any profile, for any value of the light wavelength $\lambda$. In Sect. 4 we extend the method to double-reflection systems, like the Wolter-I frequently adopted in X-ray telescopes, and in Sect. 5 we recall some experimental validation by comparison of the predictions with the measured PSFs at the SPring-8 synchrotron light source[19] and at the FERMI@Elettra FEL[20].

## 2. THE HUYGENS-FRESNEL PRINCIPLE IN 1 DIMENSION

An application of the Huygens-Fresnel principle to a 2D mirror surface can be a relevant computational load. For this reason, most wavefront propagation codes operate on mirrors known analytically or mapped at a resolution of a few millimeters. This is an acceptable sampling for mirrors operating in the visible range, in which the achieved level of surface polishing is known to reduce the optical scattering to negligible levels.

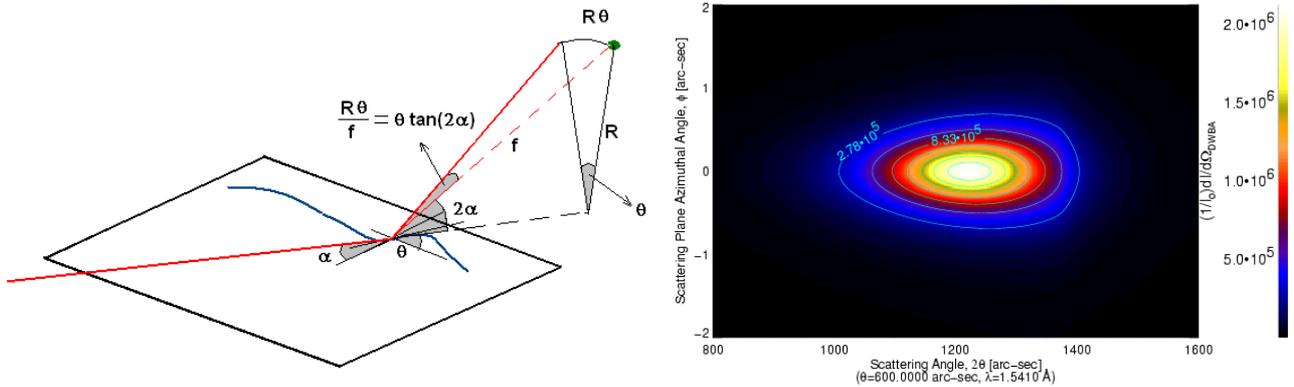

Fig. 2: justification of the 1D approximation in geometrical optics and in the scattering theory. (left) effect of a slope error $\theta$ in a grazing-incidence mirror (incidence angle $\alpha$) in the transverse (azimuthal) direction. The effective angular departure from the ideal focus (green dot) is $\theta \tan(2\alpha)$ Vs. a $2\theta$ angular deviation that the ray would experience if the slope error were in the longitudinal direction. (right) 2D computed scattering pattern off a Silicon surface for 8 keV X-rays impinging at 600 arcsec off-surface (simulated via the IMD package[22] by D. Windt**). The isophotes are ellipses extremely elongated in the incidence plane (the horizontal axis: the vertical scale is heavily stretched).

---
** http://www.rxollc.com/idl/

For X-ray mirrors, the situation clearly changes: a PSF must account for the presence of surface roughness, therefore the mirror map has to be sampled with a sub-micron lateral resolution: the resulting matrix is often too large to return a PSF computation in a reasonable time. There is a possible way out of this problem: we may work in *far-field* approximation, i.e. in adopting the Fraunhofer diffraction theory instead of the Fresnel one (as done by several authors[14], [21]). X-ray mirrors *usually* have focal lengths that, owing to the shallow angles at play, are very long with respect to their dimensions. If this can be done, the mirror diffraction pattern simply becomes the power spectrum of the mirror's Complex Pupil Function (see, e.g.[13]), which can be computed via effective numerical tools like the Fast Fourier Transform.

There is another problem in this approach, indeed. X-ray optical systems do *not* always operate in far-field conditions: for example, subsequent reflections can occur at short distances, at least comparable with the linear sizes of the mirrors. An example is provided exactly by the Wolter-I design[2], widespread in X-ray telescopes, in which the primary (parabolic) and the secondary (hyperbolic) segment are in close contact or separated - at most - by a few centimeters gap. In this case, we simply cannot compute the diffracted field by the primary mirror onto the secondary in far-field conditions, so we have to find out another method to deal with the PSF computation.

Fortunately, we can find a solution simply remarking that in grazing incidence setup the ray deviations out of the incidence plane are almost always negligible (Fig. 2, left), excepting mirrors with marked anisotropies or out-of-roundness errors, seldom met in practice. Since the off-axis angles are always small, the incidence plane is very close to the meridional plane of the mirror, so the overwhelmingly dominating contribution to the PSF broadening comes from the longitudinal (axial) errors rather than from the azimuthal (sagittal) ones. For this reason, mirror manufacturers are much more concerned by the former class of errors than by the latter. Their concern is justified, however, not only from the viewpoint of geometrical optics. Wave optics effects in X-ray mirrors, like scattering, exhibit exactly the same behavior because the angular spread is concentrated in the incidence plane (Fig. 2, right) and almost completely dependent on the rough topography in the sole axial direction. Finally, in grazing incidence the mirror's entrance pupil is a thin annulus, so that also the aperture diffraction (not always negligible in X-rays as one might think… see Fig. 10) is in practice affected by the sole annulus amplitude in the radial direction (Fig. 3).

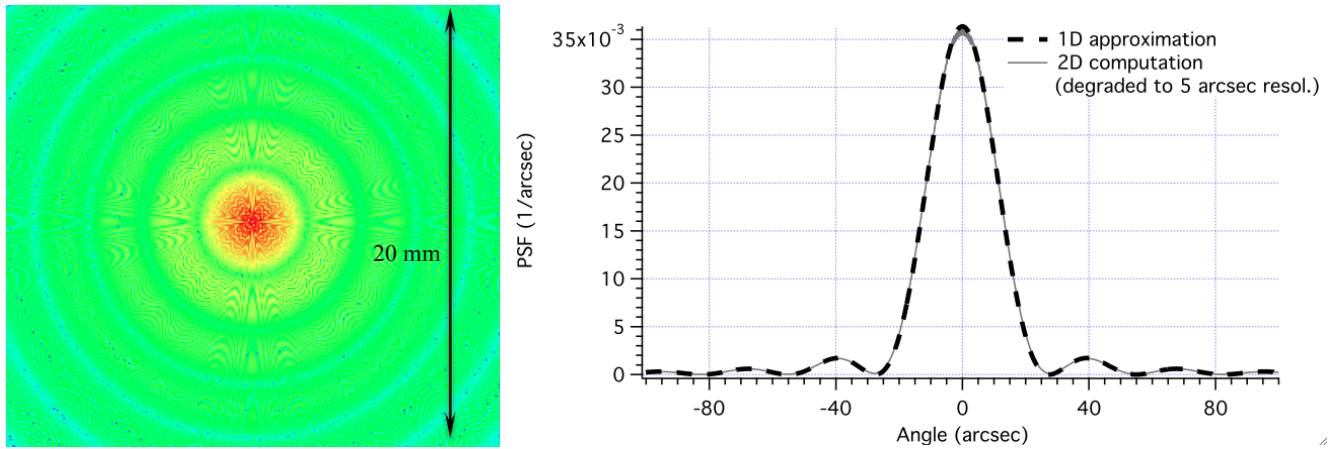

Fig. 3: justification of the 1D approximation in the aperture diffraction. (left) the 2D diffraction pattern computed for a perfect parabolic mirror with a minimum radius of 150 mm, a mirror length of 300 mm, and a focal length of 10 m at $\lambda$ = 3000 μm, in log color scale. The pupil aperture is a thin ring of 2.25 mm radial aperture. (right) integration of the 2D pattern in circular coronae returns the gray line. Exactly the same PSF is computed, but much more easily, as 1D diffraction pattern of a long straight slit of same width (dashed line).

The examples reported above suggest that in grazing incidence we can easily neglect 1) the mirror defects in the azimuthal direction and 2) the transverse extent of the PSF (a similar reasoning should also persuade the reader that in grazing incidence also the polarization states of the X-rays remain almost unchanged and that 3) we can work in scalar approximation). We have so performed the integration of the Fresnel diffraction formula in the azimuthal direction, obtaining a versatile result (Eq. 3 in the next section), suitable to compute the 1D PSF from any 1D profile of a single reflection mirror. In this way, we just have to numerically solve a 1D integral for every point of a radial line on the detector plane, not throughout all of its surface, and the computational complexity is reduced by *orders of magnitude* without being limited to the far-field configuration. Should more longitudinal profiles be measured, the PSF from each profile can be computed independently and the PSFs ensemble can be averaged to return *the* mirror PSF.

# 3. THE PSF OF SINGLE REFLECTION MIRRORS

## 3.1. PSF integral formula for isotropic sources

For the PSF computation of a focusing, grazing incidence mirror, we hereafter recapitulate some results treated in detail in[15], omitting the mathematical derivations, and adopting Fig. 4 as a reference frame. The diffracting object is a sector of an axially-symmetric mirror of length $L_1$, entirely characterized in profile by the function $x_1(z_1)$, which encompasses the nominal shape and profile errors of any spatial frequency. Anticipating a formalism extension to multiple reflections, the "1" subscript is useful to remind that this is the first reflection encountered by the X-ray beam. The X-ray source is assumed to be distant, isotropic, and point-like on the positive side of the $z$-axis, which is also the symmetry axis of the mirror. The source is located at the coordinate $z = S$, where $S$, if positive, can be either infinite like in the case of an astronomical source, or finite, although very large: the latter is the setup of on-ground X-ray testing facilities like PANTER[23] (synchrotron and FEL light also have a source at finite distance, but their emission is highly anisotropic and will be considered in Sect. 3.3). We also admit the possibility of $S < 0$ to include the case of a spherical wave converging to $z = S$. For a given location of the X-ray source, the mirror has the best focus at $x = 0$, $z = 0$, and we say $f$ to be the $z$-coordinate of the mirror's exit pupil. The source-mirror separation (with sign) is $D = S-f$. In this way, the mirror spans over the $z$-axis from $f$ to $f+L_1$. We also require that $|S| \gg f > L_1$, a condition always met in practical cases, and finally denote with $R_0$ and $R_M$ the azimuthal radii at the exit and at the entrance pupil, respectively.

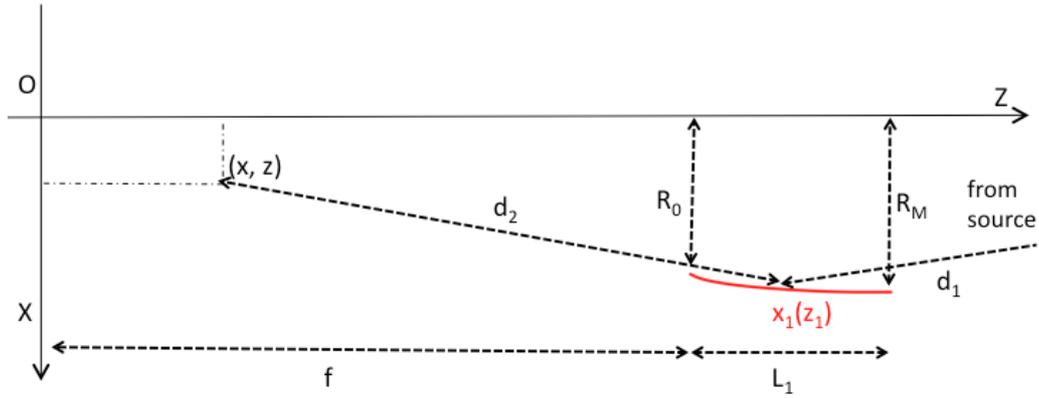

Fig. 4: the reference frame used for computing the field diffracted by a focusing mirror. The source is assumed on axis at $z$ = S.

Owing to the small curvature of the axial profile, the incidence angle for a source at infinite distance $\alpha_0$ is practically constant throughout the profile length, and so is the beam divergence $\delta \approx R_0/D$ (same sign of $D$). A simple geometric construction then shows that the true incidence angle is $\alpha_1 = \alpha_0 + \delta$ and the radial aperture of the mirror is $\Delta R_1 \approx L_1 \tan\alpha_1$, as seen from the source. We now assume the electric field at the mirror entrance to have uniform amplitude $E_0$. An integration of the Fresnel integral along the azimuthal direction[15] provides the expression of the diffracted electric field at any location in the $xz$ plane:

$$E(x,z) = \frac{E_0 \, \Delta R_1}{L_1 \sqrt{\lambda x}} \int_f^{f+L_1} \sqrt{\frac{x_1}{\bar{d}_2}} e^{-\frac{2\pi i}{\lambda}\left[\bar{d}_2 - z_1 + \frac{x_1^2}{2(S-z_1)}\right]} dz_1 \qquad (1)$$

where we omitted for simplicity the explicit dependence of $x_1$ on $z_1$, and we have set

$$\bar{d}_2 = \sqrt{(x_1 - x)^2 + (z_1 - z)^2}. \qquad (2)$$

Evaluating the field intensity in the nominal focal plane ($z = 0$) and normalizing to the collected beam intensity we obtain *a general expression for the PSF* [15],[16]:

$$\text{PSF}(x) = \frac{\Delta R_1}{L_1^2 \lambda R_0} \left| \int_f^{f+L_1} \sqrt{\frac{x_1}{\bar{d}_{2,0}}} e^{-\frac{2\pi i}{\lambda}\left[\bar{d}_{2,0} - z_1 + \frac{x_1^2}{2(S-z_1)}\right]} dz_1 \right|^2 \qquad (3)$$

where $d_{2,0}$ is Eq. 2 evaluated in $z = 0$. The computation of the integral in Eq. 3 should be in general performed numerically, after substituting the function $x_1 = x_1(z_1)$.

It is interesting to note that, to within the numerical computation accuracy[15],

$$\int_{-\infty}^{+\infty} \text{PSF}(x)\,dx = 1 \qquad (4)$$

i.e., the PSF integrated over $x$ correctly returns the entire collected intensity, for any value of $\lambda$. However, if the integration is performed over a 1D detector (a "focal line") of side $\rho$, then the integration may return a value smaller than 1. Since the PSF is absolutely normalized to the power impinging onto the mirror, a PSF normalization smaller than 1 physically indicates that the PSF is broad enough for its wings to fall out of the detector area. The FWHM computation from the PSF is straightforward. The HEW is simply computed as twice the median value of the PSF, provided that the integral of the PSF within the focal line length is larger than 50%. If this is not possible, the HEW can be computed after re-normalizing to 1 the PSF computed within the focal line extent.

A numerical computation of the integral in Eq. 3 entails a sampling of the profile $x_1(z_1)$, with $f < z_1 < f + L_1$, and of the focal line, at $-\rho/2 < x < +\rho/2$. The 1st order scattering theory allows us to determine the *minimum* sampling steps[16],[17] necessary to avoid "ghost" features in the PSF:

$$\Delta z_1 = \frac{\lambda f}{4\pi \sin \alpha_1 \, \rho} \qquad (5)$$

$$\Delta x = \frac{\lambda f}{2\pi \sin \alpha_1 \, L_1} \qquad (6)$$

where $\alpha_1 = \alpha_0 + \delta$. We notice that the number of points in the sampled profile and in the sampled focal line is the same number, $N$, and that the substitution of reasonable values in Eqs. 5 and 6 (e.g., $\lambda = 12$ Å, $f = 10$ m, $\alpha_1 = 0.5$ deg, $L_1 = 200$ mm, $\rho = 20$ mm) yields $N \approx 4 \times 10^4$, a number of terms that can be easily managed. A simple IDL code developed to implement the formulae reported in this paper, named WISE (*Wavefront propagatIon Simulation codE*), run in a computer with a 2.4 GHz processor, achieves the computation with the parameters listed above in a 5 min time. The number of required operations increases with $\lambda^{-2}$, and so does the computation time. We finally mention that this formalism can be easily adapted to extended X-ray sources, spatially and temporally coherent or not[15].

## 3.2. The PSF in the far-field limit

In the astronomical case $S \to +\infty$, and if $f \gg L_1$, then the square root in the integrand of Eq. 3 varies slowly with respect to exponential and can be approximated by the constant $(R_0/f)^{1/2}$. So Eq. 3 takes a simpler form, already derived in a previous work[16]:

$$\text{PSF}(x) = \frac{\Delta R_1}{L_1^2 \lambda f} \left| \int_f^{f+L_1} e^{-\frac{2\pi i}{\lambda}\left[\sqrt{(x_1-x)^2 + z_1^2} - z_1\right]} dz_1 \right|^2 \qquad (7)$$

The real mirror profile $x_1$ can be decomposed into the nominal profile, $x_{1n}$ (parabola, hyperbola, ellipse…) and a profile error term $x_{1e}$, including profile defects over all spatial scales. Substituting $x_1 = x_{1n} + x_{1e}$ into Eq. 7, developing the exponent and the root at the first order, and imposing that $x_{1n}$ focuses exactly to $(0,0)$, after some handling[15] we obtain the well-known, far-field approximate form of the PSF formula,

$$\text{PSF}(\varphi) = \frac{1}{\Delta R_1 \lambda} \left| \int_0^{+\infty} e^{-\frac{2\pi i}{\lambda} x_{1n} \varphi} \, \text{CPF}(x_{1n})\, dx_{1n} \right|^2 \qquad (8)$$

where we have set $\varphi = x/f$, the angular deviation from the ideal focus (the origin of the reference frame), and the CPF is the Complex Pupil Function,

$$\text{CPF}(x_{1n}) = \exp\left(-\frac{2\pi i}{\lambda} 2 x_{1e} \sin \alpha_0\right) \qquad (9)$$

We therefore see that the PSF formula (Eq. 3) correctly reduces in the far-field limit to the squared module of the CPF Fourier Transform. This particular form cannot be used, however, to compute the PSF of Wolter-I mirrors, because - among other things – in this case the square root in Eq. 1 cannot be approximated by a constant (Sect. 4).

### 3.3. PSF formula for anisotropic sources (Synchrotrons and Free Electron Lasers)

Unlike natural X-ray sources (either astronomical or bremsstrahlung tubes), the brightness distribution of synchrotrons and FELs is highly *directional and anisotropic*. A FEL like FERMI[5] in its fundamental mode propagates in spherical waves, but the intensity over the wavefront is non-uniform, with a typical Gaussian intensity distribution[24]. The electric field amplitude over a focusing mirror surface (typically, with an elliptical profile) can be written as

$$u(x_1, z_1) = \sqrt{\frac{\Delta R_1}{w}} \sqrt{\frac{2}{\pi}} \exp\left[-\frac{(x_1 - R_c)^2}{w^2}\right] \tag{10}$$

where $R_c = (R_0 + R_M)/2$, and the beam width $w$ evolves along the propagation in inverse proportion to the source width $w_0$. At the mirror location we can write, to a good approximation, $w \approx \lambda D / f w_0$. The multiplicative constant in Eq. 10 is chosen to normalize to 1 the average beam intensity.

Because the wavefront are still spherical, the PSF equation (Eq. 3) can be immediately extended to this case, weighting the integrand over the amplitude distribution:

$$\text{PSF}(x) = \frac{\Delta R_1}{L_1^2 \lambda R_0} \left| \int_f^{f+L_1} u(x_1, z_1) \sqrt{\frac{x_1}{\bar{d}_{2,0}}} e^{-\frac{2\pi i}{\lambda}\left[\bar{d}_{2,0} - z_1 + \frac{x_1^2}{2(S-z_1)}\right]} dz_1 \right|^2 \tag{11}$$

A simple application of Eq. 11 is to a perfect ellipsoidal mirror in far-field approximation. Substituting Eq. 10 into Eq. 11, and simplifying the integrand as we did in Sect. 3.2, we remain with

$$\text{PSF}(x) = \frac{\sqrt{2/\pi}}{w \lambda f} \left| \int_{R_0}^{R_M} e^{-\left[\frac{(x_1 - R_c)^2}{w^2} + 2\frac{\pi i x_1}{\lambda f} x\right]} dx_1 \right|^2 \tag{12}$$

thus, completing the square in the exponent we obtain, after some algebra,

$$\text{PSF}(x) = \sqrt{\frac{2}{\pi}} \frac{D}{f w_0} e^{-2\left(\frac{Dx}{f w_0}\right)^2} \tag{13}$$

where we assumed that $\Delta R_1 / 2w \to +\infty$, i.e., that the mirror is large enough to collect the entire Gaussian beam. Eq. 13 is exactly the source intensity profile, de-magnified by a factor of $f/D$. We have therefore obtained a geometrical optics result by application of the Fresnel diffraction. As we will see in Sect. 4, this is a very frequent situation.

### 3.4. Treatment of roughness

For a real focusing mirror, the PSF can be computed substituting into Eq. 3 (or Eq. 11 if the source is anisotropic) the real longitudinal profile $x_1(z_1)$. As already mentioned, we can figure out that $x_1$ is composed by the nominal profile $x_{1n}$ that focuses the beam to the origin of the reference frame, and a profile error $x_{1e}$ that determines the shape and the size of the focal spot. The profile error can be measured using a profilometer over the entire length of the mirror, but real instruments have a finite spatial resolution, so the measurement will be necessarily limited in spatial bandwidth. Yet an affordable PSF computation in X-rays also requires including roughness at higher spatial frequencies, at least up to the Nyquist frequency corresponding to the minimum sampling, $\nu_{\min} = 1/2\Delta z_1$ (Eq. 5). This frequency often falls in the micron range, and measurements with this resolution and accuracy cannot realistically be extended over the entire mirror length. For this reason, roughness measurement are sampled at different locations, assumed to be representative, treated as Power Spectral Density (PSD), and averaged in order to improve the statistical significance.

The final PSD is a complete statistical characterization of surface defects distributed over different spatial frequencies, but it cannot be reversed to return the original profile. The reason is that the relative phase of the Fourier components is suppressed in the squared module operation. However, we can generate *infinite* different profiles consistent with that measured PSD, selecting the component phases at random (Fig. 5). We can therefore assume that the profile $x_1(z_1)$ can be decomposed into three terms, splitting $x_{1e}$ into a directly measured term, $x_{1\text{meas}}$, and another one reconstructed from the PSD characterization, $x_{1\text{PSD}}$. So we have

$$x_1(z_1) = x_{1n}(z_1) + x_{1\text{meas}}(z_1) + x_{1\text{PSD}}(z_1) \tag{14}$$

For a complete, but non-redundant profile characterization, the minimum spatial frequency present in the PSD should concatenate, but not overlap, to the maximum frequency present in $x_{1meas}$. Moreover, complementary instruments can be used (PSI, AFM…) to cover a wide band of spatial frequencies. We remind that the 3 contributions are treated together, using Eq. 3 or Eq. 11: the separation we wrote is simply aimed at reconstructing a mirror profile with all the spectral components needed to calculate the PSF. In a sense, the separation between "figure" and "roughness" is no longer related to the different physical model adopted, but only to the "deterministic" or "statistical" measurement method. A natural question now arises: how can we make sure that the computed PSF will not depend on the particular realization of the rough profile, i.e., on the choice of the phases? There is no answer in general. Fortunately, one of the results of the 1st order scattering is that, if the smooth-surface limit is met,

$$\int x_{1PSD} dz_1 < \left(\frac{\lambda}{4\pi \sin \alpha_1}\right)^2 \tag{15}$$

then the scattering distribution *will depend only on the PSD*, and not on the particular profile. If Eq. 15 is not fulfilled, however, this is no longer guaranteed and the profile should be reconstructed in a more reliable way. We will study this problem in a subsequent paper.

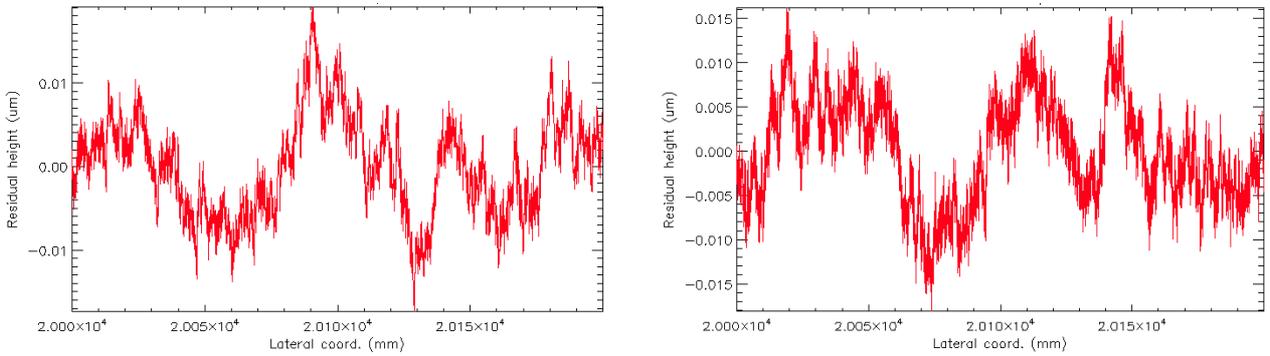

Fig. 5: reconstructed profiles from a power-law PSD with parameters $n$ = 1.5 and $K_n$ = 30 nm$^3$ μm$^{-1.5}$. In both cases, the component phases were selected at random, but the profile sampling is different. (left) at 0.25 keV, $\lambda$ = 49.6 Å, Eq. 5 returns a minimum sampling step $\Delta z_1$ = 31 μm. (right) at 1.5 keV, $\lambda$ = 8.3 Å, and $\Delta z_1$ = 5 μm. The higher content of high-frequency components is apparent.

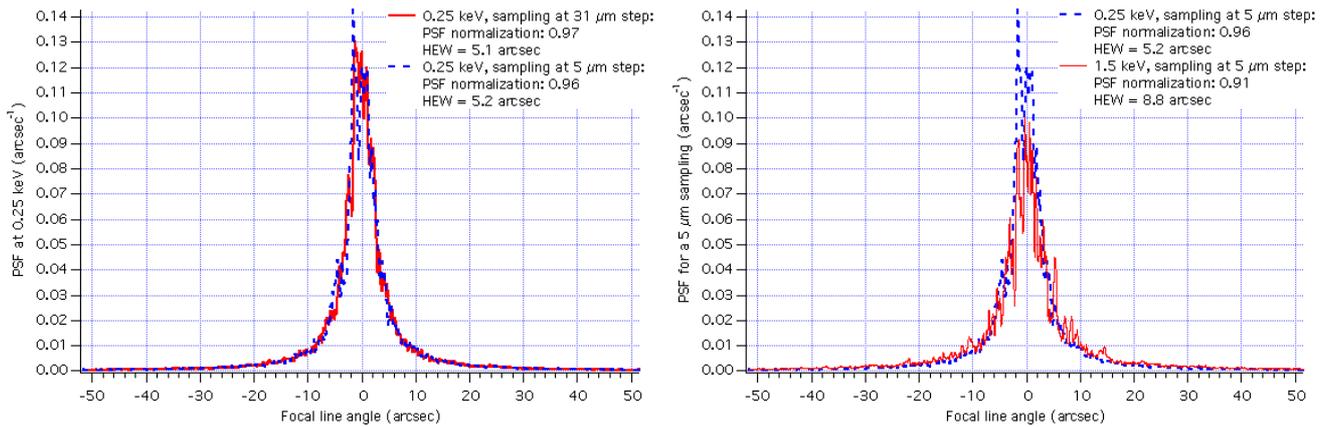

Fig. 6: PSFs computed for a geometric profile error (Eq. 24) with HEW = 5 arcsec and a superimposed roughness profile. (Left) at 0.25 keV, with two different profile samplings: the solid line PSF was calculated with the minimum sampling provided by Eq. 5, i.e. reconstructing the roughness profile at a sampling step of 31 μm (Fig. 5, left). A tighter sampling to 5 μm (like in Fig. 5, right) allows adding spatial wavelengths components down to 10 μm, but without relevant changes in the PSF. (Right) PSFs computed at 0.25 keV and 1.5 keV, obtained reconstructing *both* rough profiles at $\Delta z_1$ for the 1.5 keV energy (5 μm). Even if the spatial frequency content of the two profiles is the same, the PSF at 1.5 keV is visibly broader.

We show in Fig. 5 two examples of profile reconstruction from a PSD according the widespread power-law model[7], where $n$, $K_n$ are constants taking on values depending on the surface finishing level:

$$P(\nu) = \frac{K_n}{\nu^n} \quad (16)$$

The two simulations shown in Fig. 5 differ from each other not only for the relative phases of the components, but also because at higher energies (Fig. 5, right) the sampling step provided by Eq. 5 is smaller, so the PSD is resampled up to higher frequencies and the profile appears much more jagged. In other words, higher energies require a tighter sampling to fully compute the PSF within the detector field. The simulation at high energy will be affected by a larger amount of scattering, but this is not simply caused by the larger jaggedness of the profile. In fact, the high frequency band [60 µm – 10 µm] is *almost unseen* in the PSF computed at 0.25 keV. This is shown in Fig. 6, where we see that increasing the profile sampling beyond Eq. 5 does not sensitively change the PSF at 0.25 keV. In contrast, rough profiles with the same sampling step of 5 µm yield completely different results at 0.25 keV and 1.5 keV (Fig. 6, right). We conclude that Eq. 5 provides the correct profile sampling: a further $\Delta z_1$ reduction to account for higher spatial frequencies seems to have a minor effect on the PSF.

### 3.5. Treatment of mid-frequencies

In this section we show the behavior of a profile perturbation with a spatial period of 1 cm, i.e., falling in the spatial range of mid-frequencies. As we anticipated in Sect. 1, this is the typical kind of defects that cannot be immediately classified as "figure error", to be treated with ray tracing, and "roughness", which can be studied with the scattering theory. In order to show the non-intuitive behavior of mid-frequencies, we hereby consider as a test case a sinusoidal perturbation,

$$x_{1\text{meas}}(z_1) = A \sin\left(\frac{2\pi}{T} z_1\right) \quad (17)$$

with $T = 1$ cm and $A = 0.1$ µm, superimposed to a parabolic mirror with 0.43 deg incidence angle, 10 m focal length. Applying geometrical optics, the expected PSF would be[25]

$$\text{PSF}_{\text{geom}}(\varphi) = \frac{1}{\pi}\left[\left(\frac{4\pi A}{T}\right)^2 - \varphi^2\right]^{-1/2} \quad (18)$$

where $\varphi = x/f$. But, if we compute the PSF on the focal plane at increasing energies, from the UV range to X-rays, using the same equation (Eq. 3) assuming an isotropic source at infinite distance, the results are much complicated (Fig. 7). For comparison, we have also added in color the expectations from Eq. 18.

In the UV range ($\lambda = 1000$ Å), the PSF consists almost completely of *aperture diffraction*, with the typical sinc shape. As the energy is increased ($\lambda = 200$ Å), the aperture diffraction is reduced and the PSF shrinks. At the same time ($\lambda = 100$ Å), lateral peaks corresponding to the positions predicted by the 1st order scattering theory appear gradually. At even higher energies ($\lambda = 50$ Å), higher order scattering peaks appear, and in X-rays ($\lambda = 10$ Å) they become narrower and closely spaced. So far, the PSF has been very different from the geometrical optics predictions. However, we note that the peaks become closer and closer, and that their amplitude decays rapidly outside the domain of Eq. 18. Finally, at $\lambda = 3$ Å the peaks are so close to blend together and in hard X-rays ($\lambda = 1$ Å) the PSF computed using the Fresnel diffraction merges with Eq. 18: as expected, the geometrical optics results are found applying physical optics in the limit $\lambda \to 0$. But we can also conclude that:

1. *The geometric/scattering treatment depends on $\lambda$ and could not be established a priori*. In particular, the smooth surface limit for this case is at $\lambda = 71$ Å, just before the appearance of the 2nd order peaks. However, after this limit the geometrical optics methods are not applicable.
2. *The transition from 1st order scattering ($\lambda = 100$ Å) to geometrical optics ($\lambda = 1$ Å) is extremely slow*. We have a confirmation that we cannot set a sudden boundary between geometric and wave treatment of surface defect.
3. *What we call "geometric optics" is nothing but the superposition of high order diffraction peaks,* which become a continuum when their spacing becomes smaller than the spatial resolution of the detector, or smoothed out by the finite monochromation of the radiation in use.
4. *Also geometrical optics results can be simulated via Fresnel diffraction*, just like the aperture and surface diffraction. No frontiers need to be set between spatial frequencies, but surface defects can be treated self-consistently applying the Fresnel diffraction approach.

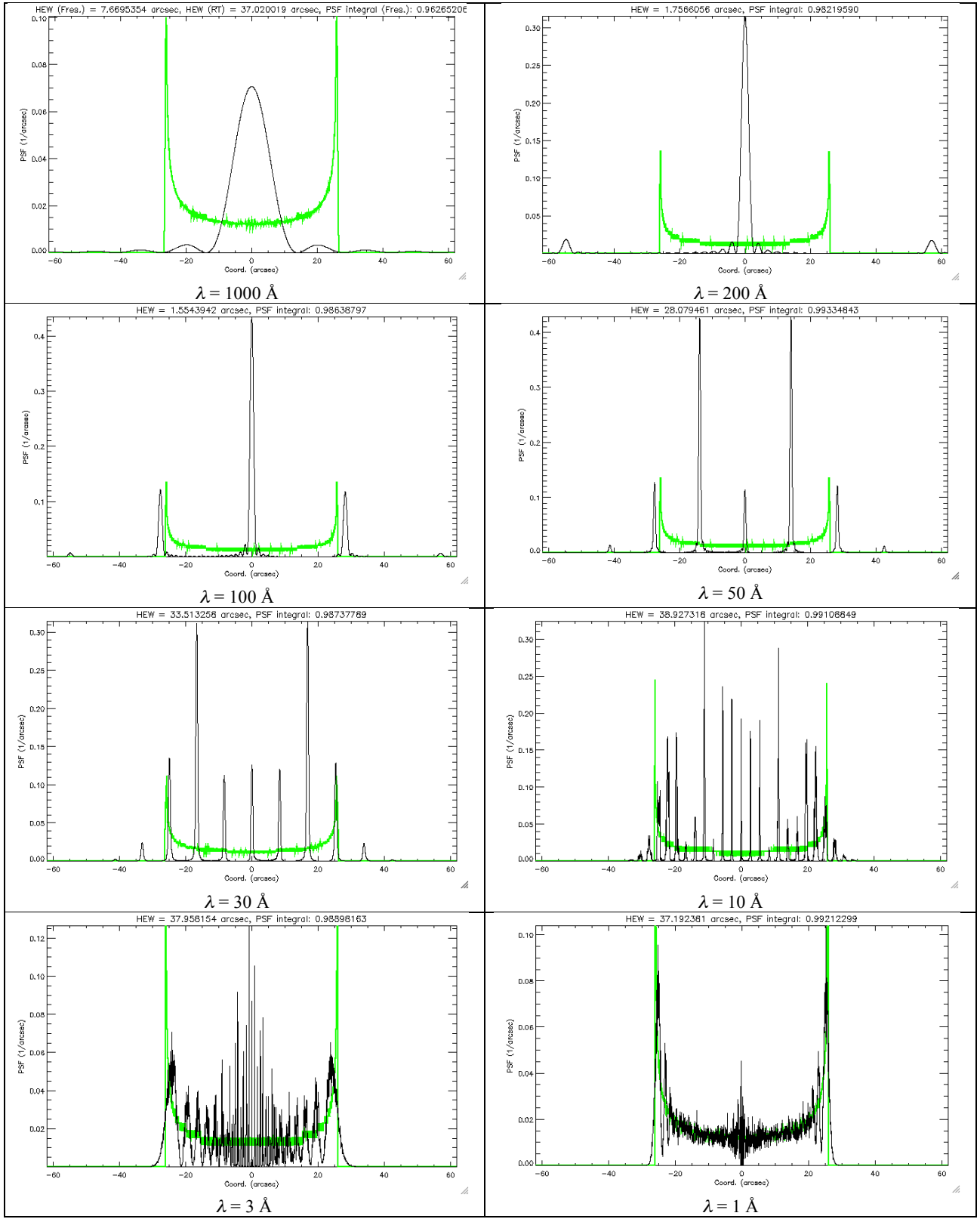

Fig. 7: the PSF of a grazing incidence ($\alpha_0$ = 0.43 deg) parabolic mirror with a *sinusoidal* perturbation (0.1 µm amplitude, 10 mm period) superimposed, for decreasing values of $\lambda$ from near UV to X-rays.

# 4. THE PSF OF DOUBLE REFLECTION MIRRORS

## 4.1. PSF integral formula for isotropic sources

The results listed in the previous sections can be extended to double reflection optical systems: this setup encompasses Wolter-I[2] or polynomial[3] mirrors used in astronomical telescopes, in which the primary and the secondary mirror reflect X-rays consecutively. The reference frame used for the calculation, shown in Fig. 8, is an extension of the scheme in Fig. 4. The primary mirror longitudinal profile is still described by the function $x_1(z_1)$, and the secondary profile (of length $L_2$) is described by the function $x_2(z_2)$, as usual including the nominal profile (a hyperbola in the Wolter-I case) and profile errors of any spatial frequency. The distance from the focal plane ($z = 0$) to the primary/secondary intersection is still denoted with $f$.

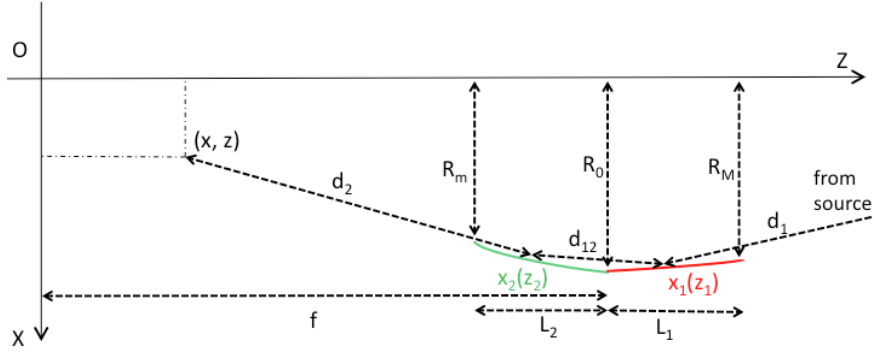

Fig. 8: the reference frame used for computing the field diffracted by a Wolter-I focusing mirror.

The computation of the PSF is done in two steps: firstly, the electric field diffracted by the primary on the secondary mirror is calculated. Eq. 1 provides the field expression:

$$E_2(x_2, z_2) = \frac{E_0 \Delta R_1}{L_1 \sqrt{\lambda x_2}} \int_f^{f+L_1} \sqrt{\frac{x_1}{\bar{d}_{12}}} e^{-\frac{2\pi i}{\lambda}\left[\bar{d}_{12}-z_1+\frac{x_1^2}{2(S-z_1)}\right]} dz_1 \qquad (19)$$

where $d_{12}$ is the distance in the $xz$ plane from a generic point of the primary mirror to a generic point on the secondary mirror:

$$\bar{d}_{12} = \sqrt{(x_1 - x_2)^2 + (z_1 - z_2)^2}. \qquad (20)$$

Because of the close spacing of the two segments, $d_{12}$ varies from $\sim L_1+L_2$ to near zero, therefore Eq. 19 cannot be reduced to the far-field form. The second step consists of calculating the PSF in focus[15], normalized to the intensity collected by the aperture pupil, which has effective radial aperture $\Delta R_{min} = \min(\Delta R_1, \Delta R_2)$, with $\Delta R_2 = L_2 \tan(\alpha_0-\delta)$.

$$\text{PSF}_2(x) = \frac{(\Delta R_2)^2}{\Delta R_{min} E_0^2 \lambda R_0 L_2^2} \left| \int_{f-L_2}^{f} E_2(x_2, z_2) \sqrt{\frac{x_2}{\bar{d}_{2,0}}} e^{-\frac{2\pi i}{\lambda}\bar{d}_{2,0}} dz_2 \right|^2 \qquad (21)$$

In Eq. 21, the *complex* function $E_2(x_2, z_2)$ contains all the relevant phase information, therefore the exponential in the integrand only depends on the distance to the focal plane:

$$\bar{d}_{2,0} = \sqrt{(x - x_2)^2 + z_2^2}. \qquad (22)$$

Exactly like Eq. 3, the PSF in Eq. 21 is independent of $E_0$ and normalized to 1 when integrated over the $x$-axis. Also for Eq. 19 and 21 we have to set appropriate sampling rules[15] for the primary mirror profile, the secondary mirror profile, and the focal line, as we already did in Sect. 3.1:

$$\Delta z_1 = \frac{\lambda}{8\pi\, \alpha_0 \sin \alpha_1}\left(1 + \frac{L_1}{L_2}\right) \qquad \Delta z_2 = \frac{\lambda f}{4\pi \sin \alpha_2 \rho} \qquad \Delta x_1 = \frac{\lambda f}{2\pi \sin \alpha_2\, L_2} \qquad (23)$$

where $\alpha_2 = \alpha_0 - \delta$. Finally, we note that the method can be extended easily to the case of anisotropic sources, in an analogous way to what we did in Sect. 3.3.

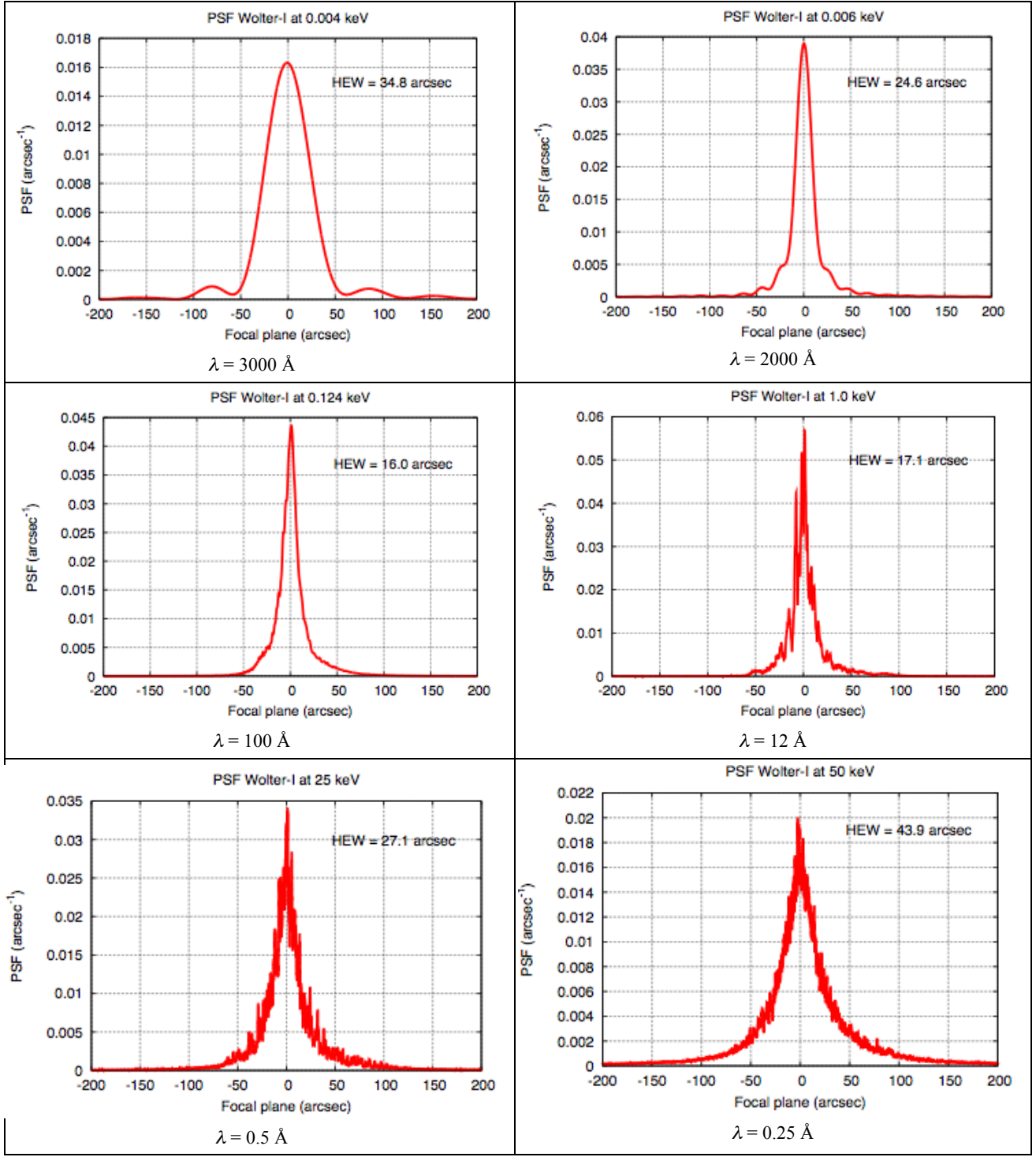

Fig. 9: simulated PSF evolution in a Wolter-I mirror profile (Eqs. 19 and 21) with a long-profile deformation aimed at a Lorentzian geometrical PSF[25], superimposed to a rough profile simulated from a power-law PSD (Eq. 16) with $n = 1.8$ and $K_n = 2.2$ nm$^3\mu$m$^{-1.8}$. $L_1 = L_2 = 300$ mm, $R_0 = 150$ mm, $f = 10$ m, and $D = +\infty$. The simulation starts from the UV range (top), dominated by aperture diffraction and almost indistinguishable from a perfect mirror. Decreasing the $\lambda$ value reduces the amount of diffraction, but the profile error effect starts to appear (middle). In soft X-rays the PSF reproduces the expectations from geometrical optics, while the roughness impact emerges when $\lambda < 10$ Å. In hard X-rays (bottom) the scattering effect has become dominant over the deterministic deformation.

## 4.2. PSF computation from UV to X-rays

In Fig. 9 we show an application of the formulae for Wolter-I mirrors. We have implemented Eqs. 19 and 21 into the WISE code (Sect. 3.1) to simulate the PSF of a Wolter-I profile adopting as $x_{\text{meas}}(z)$ for both mirror segments, of the same length $L$, an analytical shape tailored to return a Lorentzian geometric PSF[25]:

$$x_{\text{meas}}(z) = \frac{L\omega}{4\pi} \log \cos\left(\frac{\gamma z}{L}\right) \quad (24)$$

where $-L/2 < z < +L/2$ is a local coordinate, $\gamma$ is a constant slightly less than $\pi$ to avoid profile divergence at the edges and $\omega$ is the single reflection HEW. For the simulations, we have selected the parameter values listed in the caption of Fig. 9 and $\omega = 8$ arcsec for both segments. Also roughness profiles, $x_{1\text{PSD}}$ and $x_{2\text{PSD}}$, were modeled from a power-law PSD (Sect. 3.4) and added to $x_1$ and $x_2$ as per Eq. 14. Exactly like in Sect. 3.5 for the case of a mid-frequency perturbation, we have performed the computation for decreasing values of $\lambda$, *always applying the same equations* (Eqs. 19 and 21), with profiles and the focal line sampling reported in Eq. 23. Also in this case, we observe in Fig. 9 a PSF dominated by aperture diffraction in UV light, but then the deterministic deformation (Eq. 24) takes over and finally the scattering prevails, with an increasing broadening of the PSF. The interesting aspect is that the relative weights of aperture diffraction, mirror geometry, and scattering are automatically accounted for in the computation. We did not need to either separate different spectral ranges to be treated with dedicated methods, or to combine the respective PSFs via convolution or other numerical methods. The method we adopted is fully self-consistent.

## 5. EXPERIMENTAL VALIDATIONS

In this page and in the next one we briefly mention two experimental validations of the Fresnel diffraction prediction implemented in the WISE code, already treated in detail elsewhere[19],[20].

The first one is a focusing test of a Kirkpatrick-Baez mirror operated at the DiProI beamline end-station, at the FERMI@Elettra FEL. The K-B mirrors are highly polished-figured slabs in fused silica, normally endowed with the proper elliptical shape by means of mechanical benders clamped at their ends. The system is usually so effective to yield a focal spot size of 9 μm × 10 μm FWHM at $\lambda = 37.5$ nm, not far from the diffraction limit (4.1 μm × 5.9 μm FWHM). However, in this case, some profile degradation had changed the mirror response to the benders, as it could be seen in the profiles measured with the LTP. Consequently, the WISE simulation from the LTP profiles and the roughness PSD predicted a sensitive degradation (Fig. 10) of the PSF, characterized by the appearance of multiple, pronounced diffraction peaks. The focal spot was experimentally recorded by exposing a PMMA sample in the X-ray best focus, and the ablation crater was subsequently observed in a microscope: the simulated and the experimental peak positions match perfectly at both $\lambda = 20.5$ nm and $\lambda = 6.7$ nm. Also the peak intensities are in good qualitative agreement. We explicitly point out that at these values of $\lambda$ and this focusing accuracy, aperture diffraction and mirror deformation effects are almost impossible to disentangle: the PSF simulation has to be done self-consistently, like in the framework of the Fresnel diffraction theory used in this work. The problem with the K-B mirrors was later fixed completely[20].

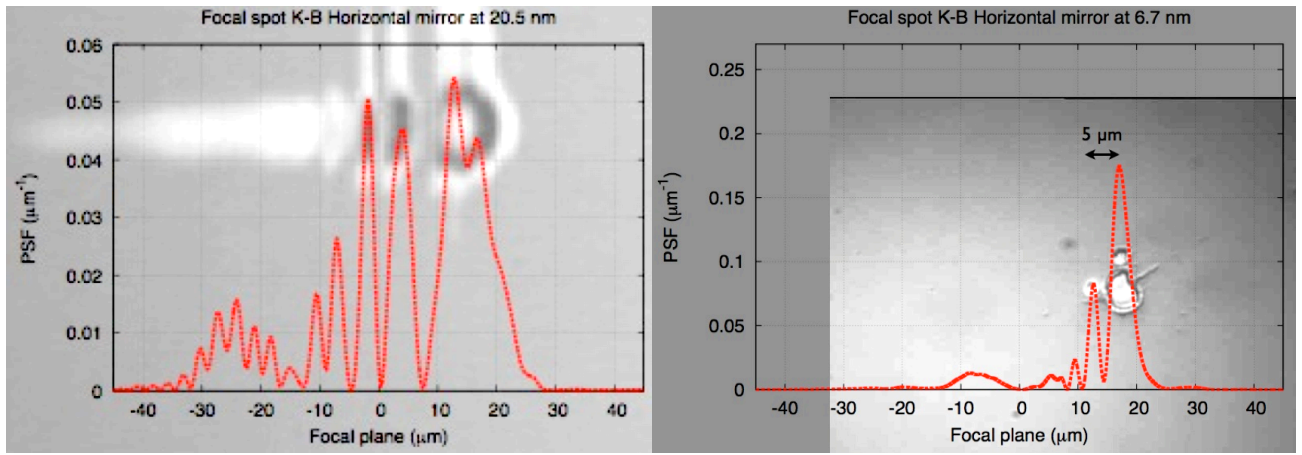

Fig. 10: PMMA ablation imprints at the DiProI beamline of FERMI@Elettra, compared with WISE predictions from profile and roughness measurements (after[20]). (Left) FEL1 at $\lambda = 20.5$ nm. (Right) FEL2 at $\lambda = 6.7$ nm. We note the excellent correspondence between the predicted PSFs (lines) and the intensity imprints (pictures).

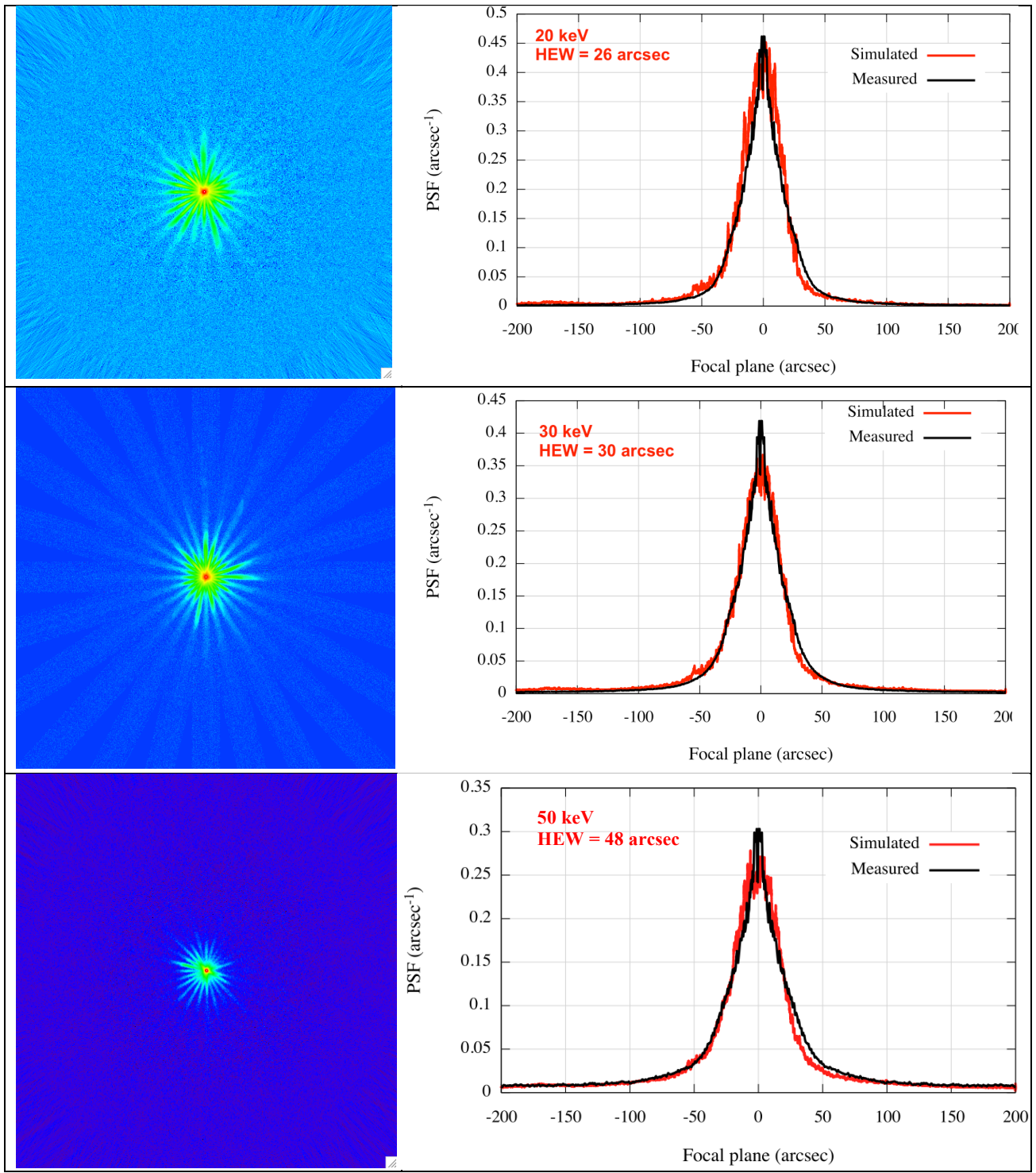

Fig. 11: (Left column) the reconstructed focal spot of the mirror shell tested at SPring-8, at (top) 20 keV, (middle) 30 keV, and (bottom) 50 keV. The focal spot *seems* to shrink because of the decreasing reflectivity that causes the focused beam to be less prominent in the background. (Right column) measured Vs. simulated PSF from measured profiles and roughness, using the WISE code. The experimental PSFs at 20, 30, and 50 keV are reproduced accurately (after[19]).

The second experimental validation was obtained in the context of the tests performed in pencil beam setup at the BL20B2 beamline of the SPring-8 radiation facility (JASRI, Hyogo prefecture, Japan) on a Wolter-I multilayer-coated mirror shell in Nickel-Cobalt alloy[19] developed for the NHXM (New Hard X-ray Mission) telescope project. The mirror shell was measured in focus at X-ray energies above 15 keV, an energy range where replicated shells often show the influence of scattering from surface roughness. The focal spots were reconstructed (Fig. 11, left) and the PSF were directly measured up to 65 keV. The mirror shell was subsequently characterized in profile and roughness (details on the measurement setup are fully given in[19]) and the Fresnel diffraction WISE code was used to compute the PSFs at 20 keV, 30 keV and 50 keV. The comparison with the experimental PSFs shows an excellent agreement (Fig. 11, right) at all three energies considered. We notice in particular that the progressive PSF broadening due to the increasing XRS with the X-ray energy is reproduced accurately.

## 6. CONCLUSIONS

In this work we have presented a method for computing a focusing mirror PSF entirely based on physical optics, which does not require any hybridization between PSFs obtained from different spatial frequency ranges. The formalism based on the Huygens-Fresnel diffraction in one dimension allows us to provide a unified treatment of surface defects at any lateral scale, including slope errors, mid-frequency errors, and roughness, at any value of the light wavelength in use. No more boundaries between spectral frequencies need to be set, with no doubt on which Physics should be correctly adopted. The relative weight of misalignments, aperture diffraction, geometry, and scattering are already accounted for in the computation. The only approximation required is that out-of plane deviations are negligible, a condition usually met for grazing incidence mirrors of good optical quality. The simplified Fresnel integrals can be implemented in any computer code (like our WISE), yielding results in very good agreement with the experiments. The computation of a PSF from a profile is a few minutes matter in soft X-rays, even with a personal computer. Far-field approximation is not needed. The source can be at finite distance or infinity, coherent or incoherent. Multiple reflections are also very easy to implement, increasing the computation time in direct proportion with the number of the reflections.

Work is in progress to extend the computation to 2D focal spot simulations of real mirrors (Fig. 12), without an excessive increase of the computation intensiveness.

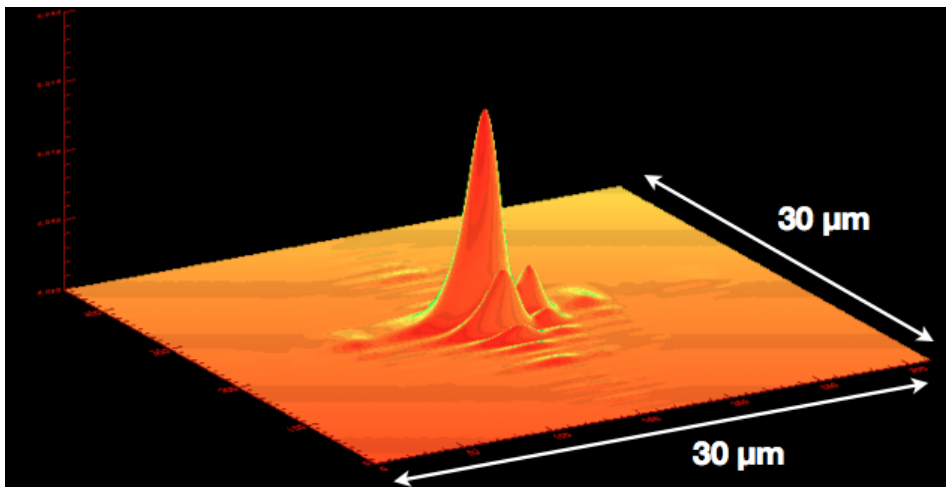

Fig. 12: a 2D focal spot simulation with WISE at the DiProI beamline at FERMI, $\lambda = 37.2$ nm, obtained from the best profiles that could be achieved with the mechanical benders, as measured with the LTP (after[20]).

## ACKNOWLEDGMENTS

Manuel Sanchez Del Rio (ESRF, Grenoble, France) and Oleg Chubar (BNL, Upton, NY) are warmly acknowledged for the expressions of interest in this work and for encouraging us to present it.